\documentclass[reprint,amsmath,amssymb,aps,floatfix]{revtex4-1}

\usepackage{epstopdf}
\usepackage{graphicx}
\usepackage{siunitx}
\usepackage{amsmath}
\usepackage{makecell}
\usepackage{subfigure}

\begin{document}
\title{The Role of Buffer Gas in Shaping the D1 Line Spectrum of Potassium Vapour}

\author{Sharaa A. Alqarni, Danielle Pizzey, Steven A Wrathmall, and Ifan G Hughes}

\affiliation{Department of Physics, Durham University, South Road, Durham, DH1 3LE, UK}

\affiliation{Corresponding author: sharaa.alqarni@durham.ac.uk}

\begin{abstract}
In this study, we investigate the effect of buffer gas and magnetic field on the spectral line shapes of the potassium D1 transition using sealed vapour cells filled with varying amounts of neon as a buffer gas. Employing a dual-temperature control system, we independently manipulate the cell body and stem temperatures to explore Doppler and collisional effects on the spectrum. Our results show how the Voigt spectral profile changes from Gaussian- to Lorentzian-dominated forms due to pressure broadening and shifts caused by collisions between potassium atoms and neon. Our measurements are in excellent agreement with the literature values for potassium-neon collisions. For the first time we were able to incorporate the buffer-gas shift and broadening into the modified Voigt profile via the {\it ElecSus} code, and found excellent agreement between the predicted and measured line profiles. We also analyse the potassium D1 spectral lines in the hyperfine Paschen-Back regime using strong magnetic fields, demonstrating how Zeeman splitting modifies the pressure-broadened line shape. This work provides valuable insights into collision-induced broadening and shifts, enhancing our understanding of potassium spectroscopy and its application in the development of advanced magneto-optical filters for solar physics and other applications. 
\end{abstract}

%
%
%
\maketitle
%
%

\section{Introduction}
Thermal vapours have long served as a crucial platform for demonstrating fundamental experimental techniques in atomic physics. Notable examples include coherent population trapping~\cite{arimondo1996v}, electromagnetically induced transparency~\cite{finkelstein2023practical}, and slow light~\cite{Slowlight,boyd2008slow}. Furthermore, significant experimental breakthroughs, such as quantum memory for light~\cite{julsgaard2004experimental, reim2011single} and the creation of fluids of light~\cite{fontaine2018observation, piekarski2021measurement}, have been accomplished using these systems. Thermal vapours also find broad application in fields such as spectroscopy~\cite{burdekin2020single, haupl2025modelling}, terahertz imaging~\cite{downes2023practical}, narrowband optical filters~\cite{gerhardt2018anomalous, uhland2023build, alqarni2024device}, and magnetometry~\cite{fabricant2023build}. The inherent properties of alkali-metal atomic vapours, including the high resonant optical depth, long coherence times, and well-understood atomic interactions, simplify both experimental setups and theoretical modelling, making them particularly suitable for such studies~\cite{pizzey2022laser}. Specifically, this work examines the potassium D1 spectral line in commercially obtained vapour cells. These cells include one with no buffer gas and two others pre-filled with neon at pressures of 60 Torr and 100 Torr. This range of pressures allows for a detailed investigation of collisional broadening and shift effects.

Potassium (K) vapour, specifically at the D1 transition (4S$_{1/2}$~--~4P$_{1/2}$) at 769.9~nm, is essential for developing solar filters~\cite{hale2016performance, hale2020modelling, refId1,refId2, giovannelli2020tor}. These filters, which employ cascaded atomic vapour cells in conjunction with a telescope for solar observation ~\cite{gelly2003detection}, require a single transmission peak centred on the atomic transition~\cite{erdelyi2022solar}. This is achieved by precisely controlling parameters such as longitudinal magnetic field, vapour temperature, cell length, and pressure broadening. Theoretical modelling of these parameters is facilitated by $ElecSus$, a software package for computing the electric susceptibility of atomic media~\cite{Zentile2015b,keaveney2018elecsus}. However, prior to this work, $ElecSus$ did not fully account for buffer-gas effects. Herein, we introduce a collisional model for buffer gases into $ElecSus$, enabling accurate simulation of the pressure broadening and shift of the potassium D1 line. The inclusion of neon buffer gas is particularly crucial for tuning the filter's spectral profile by controlling these broadening and shifts effects.

The pressure broadening ($\Gamma$) and shift ($S$) in alkali-metal vapours have been extensively studied over several decades~\cite{ottinger1975broadening, romalis1997pressure, lwin1978collision, lewis1980collisional,pitz2012pressure}, typically using gas-handling systems to manipulate and analyse their pressure dependence. In contrast, our approach utilises pre-filled vapour cells, where the amount of gas is fixed at the time of sealing, allowing us to examine the absorption spectra of the potassium D1 line as a function of temperature. We investigate a specialised vapour cell designed with independent temperature control for the cell body ($T_{\mathrm{c}}$) and the stem ($T_{\mathrm{s}}$). This configuration enables us to independently evaluate the effect of $T_{\mathrm{c}}$ on Doppler broadening and $T_{\mathrm{s}}$ on vapour pressure and line depth. Moreover, we explore the impact of neon, which induces homogeneous broadening and shifts the absorption frequency due to increased collision rates. The gradual transition from Gaussian to Voigt profile---a convolution of Gaussian and Lorentzian profile---highlights the change from a thermally dominated system to one where atomic collisions define the spectral features, particularly the width and centre of the absorption line. Precise control of the atomic vapour temperature and magnetic field is paramount to the accuracy of these measurements. 

Additionally, we examine the potassium D1 spectral lines in the hyperfine Paschen-Back (HPB) regime. In this regime, a strong external magnetic field decouples the total angular momentum of the electron $\hat{F}$ into $\hat{J}$ and the nuclear spin $\hat{I}$. The quantum numbers in weak field,  $\it{F}$ and projection $m_F$, are replaced with projections $m_J$ and $m_I$ describing the splitting of the atomic levels. The magnetic field induces distinctive changes to the energy levels and their splittings, significantly affecting the absorption profile. In the HPB regime, the Zeeman effect dominates over the hyperfine splitting, causing the energy levels to shift linearly with the magnetic field, according to $\Delta E\,=\,g_J\,m_J\,\mu_{\rm B}\,B$, where $g_J$ is the Land\'{e} g-factor for $J$ and $\mu_{\rm B}$ is the Bohr magneton. Numerous recent studies have explored this regime, ranging from fundamental investigations~\cite{sargsyan2018hyperfine,mottola2023electromagnetically} to applied research~\cite{tremblay1990absorption,briscoe2023voigt,higgins2024fine}. The field strength required to reach this regime for potassium can be estimated using $B_{\rm HPB}=A_{\rm hf}/\mu_{\rm B}$, where $A_{\rm hf}$ is the magnetic dipole constant for the ground term, calculated to be around
156~G~\cite{pizzey2022laser,sargsyan2018hyperfine}. 

In this study, we explore magnetic fields in the range 700--1200 G. This work is motivated by our research group's ongoing development of solar filters and advancements of potassium spectroscopy~\cite{erdelyi2022solar}. Our primary objective is to quantify spectral line broadening and shifts due to collisions with buffer gases, particularly neon, and magnetic fields. By gaining a deeper understanding of these effects, we aim to refine the design of magneto-optical filters, enhancing solar physics applications through improved filtering precision and detection capabilities for solar observations. 

The remainder of the paper is structured as follows: Section~\ref{sec:ther_code} presents the theoretical framework for analysing the temperature dependence of spectral line broadening and shift, highlighting the role of collisional processes in determining the optical properties of potassium vapour in the presence of a buffer gas.  Section \ref{sec:exp_details} details the experimental procedures for performing potassium spectroscopy as a function of temperature. 
Section~\ref{sec:exp_results} presents experimental results on the potassium D1 line, both with and without an applied magnetic field. Finally, Section~\ref{sec:conclusion} provides conclusions and an outlook.

\section{\label{sec:ther_code}{Theoretical model}}

The theory of collisional broadening dictates that collisional effects are temperature-dependent~\cite{zameroski2011pressure, pitz2012pressure, li2016pressure, pitz2014pressure}. Consequently, the gas broadening and shift rates,  ${\gamma_{\mathrm{j}}}$ and  ${\delta_{\mathrm{j}}}$ respectively, can be expressed as functions of temperature $T_{\rm j}$ using the following equations:

\begin{equation}
{\gamma_{\mathrm{j}}}= {\gamma_{\mathrm{r}}}\left(\frac{T_{\mathrm{r}}}{T_{\mathrm{j}}}\right)^n
\label{eqn:1}
\end{equation}
and
\begin{equation}
{\delta_{\mathrm{j}}} = {\delta_{\mathrm{r}}}\left(\frac{T_{\mathrm{r}}}{T_{\mathrm{j}}}\right)^n.
\label{eqn:2}
\end{equation}
Here, ${\gamma_{\mathrm{r}}}$ and ${\delta_{\mathrm{r}}}$ represent the gas broadening and shift rates at a reference temperature ${T_{\mathrm{r}}}$, and $n$ is the temperature-dependent coefficient. For potassium D1 in neon at ${T_{\mathrm{r}}}$~=~328~K, the values of ${\gamma_{\mathrm{r}}}$ and ${\delta_{\mathrm{r}}}$ are 6.14 and -1.27~ MHz/Torr, respectively~\cite{pitz2012pressure}. The theoretical value for $n$ is predicted to be 0.5 for neon~\cite{zameroski2011pressure,pitz2012pressure,pitz2009pressure}.

\begin{figure}[htbp]
\centering
\includegraphics[width = \linewidth]{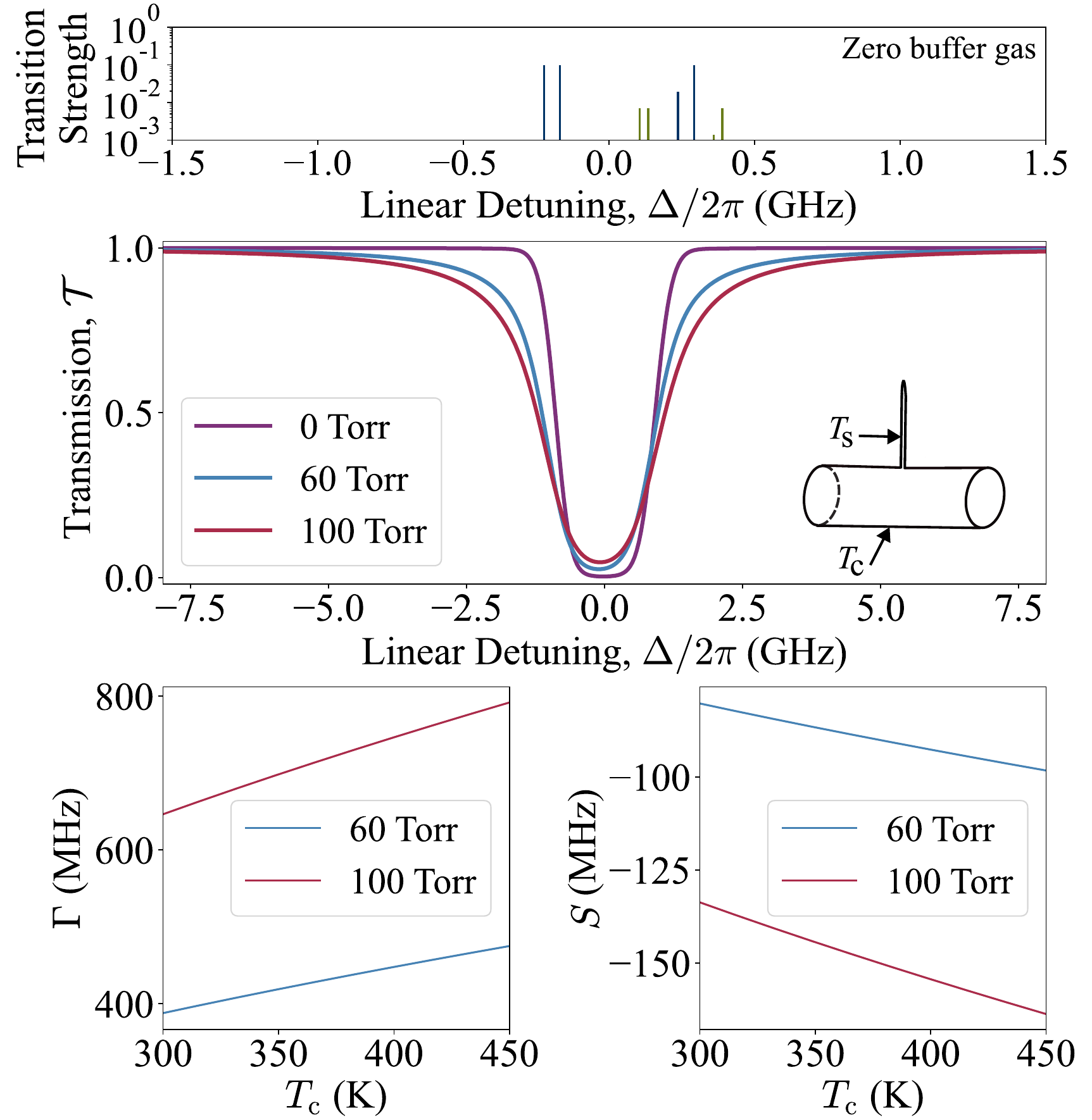}
\caption{Theoretical prediction for the D1 line in K, through a 75~mm natural abundant vapour cell. The upper panel shows the transition strengths, weighted by the
isotopic abundance~\cite{hanley2015absolute}.  Zero of the detuning axis is the weighted line-centre of the respective D1 line. The purple, blue and red spectra correspond to a vapour cell containing no buffer gas, 60 Torr, 100 Torr of neon, respectively at $T_{\mathrm{c}}$ = 363~K. A constant temperature difference of around \SI{20}K is maintained between the body of the cell and the reservoir. This temperature gradient prevents K from condensing on the windows of the cell. The inset shows a schematic of the 75 mm vapour cell used, with a central reservoir. The lower plots illustrate the dependence of buffer gas-induced broadening $T$ and spectral shift $S$ on temperature.}
\label{fig:theory k spec no B}
\end{figure}

To relate these rates to the pressure within the cell, we use the ideal gas law. The pressure ${P_{\mathrm{j}}}$ at any temperature ${T_{\mathrm{j}}}$ is:

\begin{equation}
{P_{\mathrm{j}}}= {P_{\mathrm{i}}}\left(\frac{T_{\mathrm{j}}}{T_{\mathrm{i}}}\right),
\label{eqn:3}
\end{equation}
where $P_{\mathrm{i}}$ is the initial fill pressure at an initial temperature $T_{\mathrm{i}} = 293$~K, corresponding to room temperature at which the cells were filled. 

Using the temperature dependence defined in Eqn.~\ref{eqn:3}, we can express the gas broadening, ${\Gamma_{\mathrm{j}}}$, and the shift, ${S_{\mathrm{j}}}$, as functions of the buffer gas pressure, $P_{\mathrm{j}}$: 
\begin{equation}
{\Gamma_{\mathrm{j}}}= {\gamma_{\mathrm{j}}}{P_{\mathrm{j}}}
\label{eqn:4}
\end{equation}
and
\begin{equation}
{S_{\mathrm{j}}} = {\delta_{\mathrm{j}}}{P_{\mathrm{j}}}~,
\label{eqn:5}
\end{equation}
where $\gamma_{\mathrm{j}}$ and $\delta_{\mathrm{j}}$ are the temperature-dependent broadening and shift rates from Eqns.~\ref{eqn:1} and \ref{eqn:2}, respectively, and $P_{\mathrm{j}}$ is the buffer gas pressure at temperature $T_{\mathrm{j}}$ from Eqn.~\ref{eqn:3}. 

Figure~\ref{fig:theory k spec no B} presents theoretical potassium D1 transmission spectra, derived using $ElecSus$ and Eqns.~\ref{eqn:1}-\ref{eqn:5}, for cell body and stem temperatures of 363~K and 343~K, respectively. It compares three vapour cells with fill pressures of 0, 60, and 100~Torr. The top subplot of Figure~\ref{fig:theory k spec no B} shows the transition line strengths, weighted by isotopic abundance, where the blue and green lines correspond to $^{39}$K and $^{41}$K, respectively. Note that we have neglected $^{40}$K due to its negligible natural abundance of 0.0117\%.

At 0~Torr, the Voigt lineshape is a Gaussian-dominated Doppler-broadened profile, because the Doppler width is two orders of magnitude larger than the Lorentzian contribution from natural broadening. Given that the ground state hyperfine splitting and isotope shift are less than the Doppler width, only one absorption line is seen. Notably, for the 0~Torr vapour cell, the optical transmission returns to 1 within a few GHz of the atomic reference frequency, indicating a narrow spectral feature. As the fill-pressure increases to 60 Torr, collisional effects manifest as a broadening ($\Gamma \approx 430$~MHz) and shift ($S \approx -89$~MHz), resulting in a Voigt lineshape. In contrast to the buffer-gas-free cell, for the 60~Torr and 100~Torr cells the optical transmission does not return to 1 until approximately $\pm$5~GHz, indicative of the Lorentzian component of the pressure broadening dominating the Voigt profile, especially in the wings. At a temperature of 314~K, the Doppler width is approximately 800 MHz, which is comparable to the 100 Torr pressure broadening width, highlighting the significant influence of collisions at this pressure. Furthermore, exploring temperatures above 314~K enables us to observe the regime where pressure broadening dominates over thermal (Doppler) effects. At 100~Torr, these effects intensify ($\Gamma \approx 704$~MHz, $S\,\approx\,-145$~MHz). The lower plots show $\Gamma$ and $S$ variations with temperature at 60 and 100~Torr. $\Gamma$ increases with temperature, more steeply at 100 Torr, due to higher collision rates. $S$ is consistently negative, increasing in magnitude with temperature, and is larger at 100 Torr. These results highlight the temperature and pressure dependence of spectral profile, line broadening and shift, demonstrating the impact of collisional dynamics on potassium vapour's optical properties.

\section{\label{sec:exp_details}{Experimental Details}}

\begin{figure}[htbp]
\centering
\includegraphics[width = \linewidth]{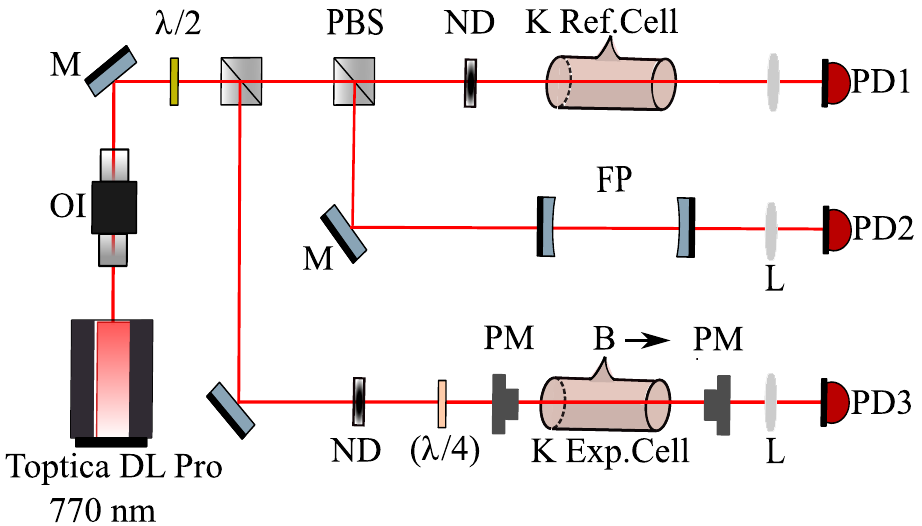}
\caption{Schematic of the experimental setup. Light from
a diode laser (DL) passes through an optical isolator (OI)
and is split into two paths using a polarising beam splitter (PBS) cube. The first path is necessary for linearising the laser scan by means of a Fabry-P\'erot (FP) etalon, while spectroscopy through a natural abundance, buffer-gas free potassium reference vapour cell provides an absolute frequency reference \cite{pizzey2022laser}. The transmission through the 100~mm length reference vapour cell (labelled ``K Ref. Cell'' in the figure) is detected on photodiode PD1, while the Fabry-P\'erot etalon is detected on photodiode PD2. The second path is for experimentation of the buffer-gas pressure within the vapour cell, the vapour cell temperature and the magnetic field the vapour cell is situated in. This vapour cell is labelled ``K Exp. Cell'' in the figure and the transmission is detected on photodiode PD3. The magnetic field is produced using axially magnetised annular permanent magnets \cite{pizzey2021tunable}, which are depicted as ``PM'', while various optics are used for beam steering, size and intensity control. Key -- M: Mirror; ND: Neutral Density filter; L: Lens; $\lambda/2$: Half Waveplate; $\lambda$/4: (optional) Quarter Waveplate.} 
\label{fig:setup}
\end{figure}

The experimental set-up, depicted in Figure~\ref{fig:setup}, utilised a laser beam resonant with the potassium D1 line. The beam was split into two paths: a reference path for laser scan linearisation and frequency calibration, and an experimental path for investigating buffer gas effects and Zeeman splitting.

The reference path, kept constant throughout the investigation, employed a Fabry-P\'erot etalon for linearising the laser scan and a 100 mm natural abundance potassium vapour cell as a frequency reference~\cite{pizzey2022laser}. The reference cell was resistively heated to achieve approximately 50\% absorption, ensuring a stable frequency reference.

The experimental vapour cells were manufactured on request by Precision Glassblowing, and constructed from fused silica tubes, featuring flat, parallel anti-reflection coated windows. To minimise potassium condensation on the windows, each cell incorporated a 40 mm stem, designed to establish a temperature differential. Cells were sealed either under vacuum (no buffer gas) or with neon gas at pressures of 60 Torr and 100 Torr at room temperature. The vapour cells were heated using two independent resistive cartridge heaters, allowing for independent control of the stem temperature ($T_{\rm s}$) and the cell body temperature ($T_{c}$). Thermocouples were attached to the stem and the cell body for monitoring.  Throughout the experiments, the cell body temperature $T_{\mathrm{c}}$ was maintained approximately \SI{20}K above the stem temperature. The optical power used throughout the investigation for the experimental vapour cell was 800~nW, with corresponding beam waists of \SI[separate-uncertainty]{2.75\pm0.02}{mm}\,$\times$\,\SI[separate-uncertainty]{2.56\pm0.02}{mm}. This ensures spectroscopy in the weak-probe regime~\cite{weakprobe}.

\subsection{Buffer gas pressure investigation}
\label{subsec:buffdetails}

This study investigated the buffer gas-induced shift ($S$) and broadening ($\Gamma$) of the potassium D1 line, examining the influence of both vapour cell temperatures and initial buffer gas pressures on these parameters. The experimental vapour cells utilised in this study were 75~mm in length.

To achieve measurable absorption of the probe beam with a satisfactory signal-to-noise ratio, the potassium vapour cells were heated to a minimum of 313~K. At this temperature, the atomic vapour density was sufficient to produce significant absorption features. The resultant Doppler width, approximately 800~MHz, significantly exceeds the ground-state hyperfine splitting of potassium isotopes ($^{39}$K: 461.7~MHz; $^{41}$K: 254.0~MHz). Consequently, the hyperfine structure remains unresolved, and the observed D1 line transmission spectrum exhibited a single Doppler-broadened profile, reflecting contributions from the ground state \cite{bruner1998frequency}.

The experimental results are shown and discussed in Section~\ref{result part a}.

\subsection{Zeeman-splitting the spectral line}
\label{subsec:zeemandetails}

This study investigated the effect of introducing a magnetic field. The magnetic field was produced using axially magnetised annular permanent magnets, similar to those described in \cite{pizzey2021tunable}, in a Faraday configuration, wherein the magnetic field was parallel to the $k$-vector. In this configuration, $\sigma^{\pm}$ transitions are driven by left- and right-hand circularly polarised light~\cite{adams2018}. As there is no component of the electric field along the direction of the magnetic field, $\pi$ transitions are not excited.

The annular magnets were designed to accommodate a vapour cell of length 25~mm (without a stem) and to exhibit an root-mean-square variation in the magnetic field along the length of the vapour cell on the order of 1-2\% at a maximum magnetic field of 1.4~kG (a typical value used in solar magneto-optical filters~\cite{refId1}). With the addition of the 6~mm cylindrical stem on the vapour cells, the maximum magnetic field achievable by the annular magnets was approximately 1.1~kG, with a root-mean-square field variation of 3\%. Rather than utilising the 75~mm length vapour cell, 25~mm vapour cells (with a stem) were employed, ensuring that all of the atomic vapour was contained within the magnetic field. The magnetic field strength was reduced by increasing the separation between the permanent magnets positioned on either side of the vapour cell stem. Three distinct magnet separations, and consequently three magnetic field strengths, were investigated for a vapour cell sealed with a neon gas pressure of 60~Torr. This variation in magnetic field strength allowed for the characterisation of the Zeeman splitting of the potassium D1 line. The experimental results are shown and discussed in Section~\ref{subsec:with B}. 

\section{Experimental Results\label{sec:exp_results}}

This section presents the experimental results obtained from spectroscopic studies of the potassium D1 line, focusing on two key aspects: the influence of buffer gas and the impact of an applied magnetic field. Firstly, we examine the effects of neon buffer gas, specifically exploring the broadening, shift, and temperature dependence observed at varying buffer gas pressures. Secondly, we investigate the evolution of the spectral line shape in the presence of a magnetic field, revealing the Zeeman splitting and its dependence on polarisation and field strength. These investigations provide valuable insights into the collisional dynamics and magnetic properties of the potassium vapour.

\subsection{\label{result part a}{Effects of buffer gas on the potassium D1 line: Broadening, shift, and temperature dependence}}

\begin{figure}[htbp]
    \centering
\includegraphics[width = \linewidth]{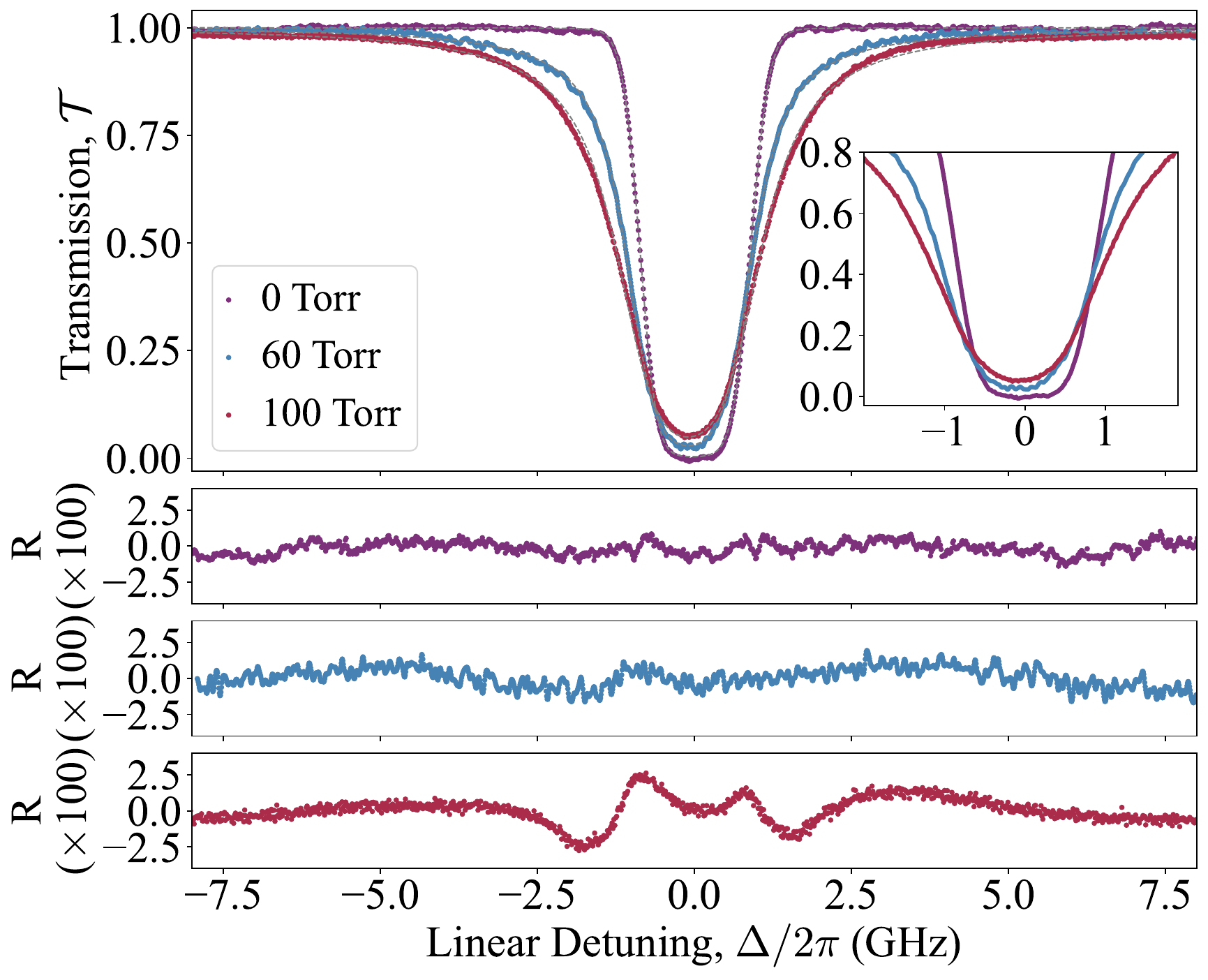}
\caption{Measured and theoretical transmission spectra of the potassium D1 line as a function of linear detuning $\Delta /2\pi$ in a 75 mm natural abundance potassium vapour cell. The solid lines represent the experimental data, while the grey dashed lines depict the theoretical spectra calculated using $ElecSus$. The spectra, from top to bottom, correspond to cells with zero buffer gas (purple), 60 Torr neon (blue), and 100 Torr neon (red), all at a cell body temperature ($T_{\rm c}$) of \SI[separate-uncertainty]{370\pm5}{K} and a stem temperature ($T_{\rm s}$) of \SI[separate-uncertainty]{350\pm5}{K}, resulting in a Doppler width of  857~MHz. The inset provides a magnified view of the central region, highlighting the spectral shifts ($S$) induced by varying buffer gas pressures. The residuals, shown below the main plot, demonstrate excellent agreement between the experimental data and the $ElecSus$ model predictions \cite{hughes2010measurements}.}. 
\label{fig:k spec}
\end{figure}

We fitted the experimental spectra using $ElecSus$, by means of a differential evolution algorithm, allowing the stem temperature, cell temperature, shift, and broadening to vary. Upper and lower bounds were imposed on $T_{\rm c}$ and $T_{\rm s}$ of $\pm$\,5\,K. Figure~\ref{fig:k spec} presents the measured transmission spectra of the potassium D1 line as a function of linear detuning for three distinct neon buffer gas filling pressures. Also shown are the $ElecSus$ fits to the data, with normalised residuals displayed underneath. All spectra were acquired at a cell body temperature ($T_{\rm c}$) of \SI[separate-uncertainty]{370\pm5}{K}, corresponding to cells with no buffer gas, 60 Torr, and 100 Torr of neon, respectively.

In the absence of buffer gas, the Voigt transmission profile is Doppler-dominated. The presence of neon buffer gas significantly broadened the transmission spectra, with the additional broadening increasing proportionally to the amount of buffer gas. For the 60 Torr cell, the linewidth ($\Gamma$) was measured to be \SI[separate-uncertainty]{405\pm3}~{MHz}, while for the 100 Torr cell, $\Gamma$ was \SI[separate-uncertainty]{798\pm3}{MHz}. This broadening arises from collisional interactions between potassium and neon atoms, which increase with pressure and temperature, contributing an additional Lorentzian component to the Voigt profile. The Gaussian component, determined by Doppler broadening, remained relatively constant.

Furthermore, the transmission spectra exhibited a pressure-dependent shift. The broadening ($\Gamma$) and shifts ($S$) extracted for the three vapour cells are listed in Table~\ref{tab:gamma-$S$ of K exp spec}. They show excellent agreement with the theoretically predicted broadening and shifts as shown in Fig.~\ref{fig:theory k spec no B}. The non-zero, positive, shift measured for the vapour cell containing no buffer gas is a systematic frequency offset, which is applicable to all datasets; this arises due to the difficulty in linearising and referencing the laser scan.

\begin{table}[]
    \centering
    \begin{tabular}{| c| c |c |}
         \hline
        ${P_i}$~(Torr) & 
        $\Gamma$~(MHz)& $S$~(MHz)\\
         \hline
        0 &0&${0.90\pm0.03}$\\
        60 &${405\pm3}$&${-87\pm2}$\\
        100 &${798\pm3}$&${-94\pm2}$\\
        \hline
    \end{tabular}
    \caption{Table of the $\Gamma$ and $S$ values for each spectrum shown in Figure \ref{fig:k spec}.}  
    \label{tab:gamma-$S$ of K exp spec}
\end{table}

We compared the experimentally determined $\Gamma$ and $S$ values (Table~\ref{tab:gamma-$S$ of K exp spec}) with theoretical simulations (Figure~\ref{fig:theory k spec no B}), based on the supplier's stated 60~Torr and 100~Torr fill pressures. To validate these initial fill pressures, we extracted the broadening and shift parameters from $ElecSus$ when we vary the cell temperature $T_{\rm c}$, and applied Eqns.\,\ref{eqn:4} and \ref{eqn:5} with the gas broadening and shift rates (from Pitz \textit{et al.} \cite{pitz2012pressure}). For each cell, spectra were taken at six different temperatures, spanning the range 370~K to 445~K. As we have noted above, for potassium the Doppler and pressure broadening exceed the atomic hyperfine splitting, and, as a consequence, one broadened feature is seen. This is in contrast to the heavier alkali metals rubidium and caesium where more than one line is seen. For potassium it is therefore more difficult to extract the individual Gaussian and Lorentzian contributions to the lineshape. For these fits we constrain the cell temperature to be that measured by the thermocouple, and also perform fits within limits of $\pm$\,5\,K of this value to allow for a possible offset between the cell and thermocouple temperatures. Historically, discrepancies of a few degrees have been noted between thermocouple readings and cell temperatures~\cite{hanley2015absolute}. The average initial fill pressures were determined to be (57.4\,$\pm$\,0.6)~Torr and (96\,$\pm$\,1)~Torr of Ne, on the assumption that the vapour cells are sealed at a temperature of 293\,K. These values are close to the expected values.

\subsection{\label{subsec:with B}{Evolution of the spectral line shape in the presence of a magnetic field}}

Following the buffer gas characterisation, we examined Zeeman splitting in the potassium D1 line. Firstly, we analysed the magnetic field dependence of Zeeman splitting using linearly polarised light. Subsequently, we investigated the influence of polarised light on the Zeeman spectra at a fixed magnetic field. The Zeeman shifting of spectral lines, and the interaction of different polarisation states of light with the atoms, are crucial components of magneto-optical filters used for solar studies~\cite{hale2020modelling, erdelyi2022solar}, and for other narrowband filtering applications~\cite{gerhardt2018anomalous}.

\subsubsection{Linear polarisation spectroscopy of Zeeman splitting at varying magnetic fields}

\begin{figure}[htbp]
\centering
\includegraphics[width=1\linewidth]{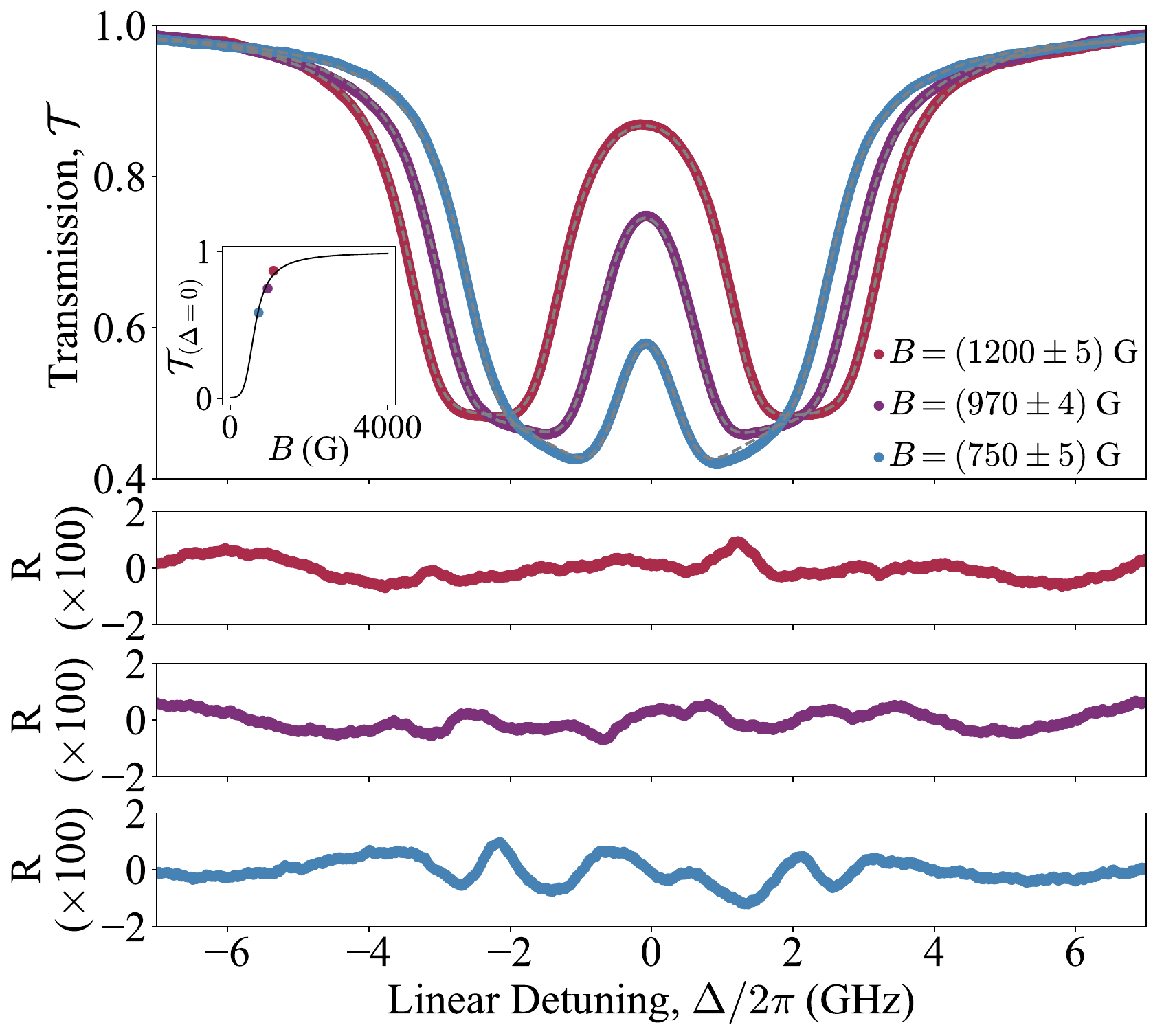}

\caption{Transmission spectra of the potassium D1 line at three magnetic field strengths $(1200 \pm 5)$~G, $(970 \pm 4)$~G, and $(750 \pm 5)$~G, measured in a 25 mm vapour cell with 60 Torr of neon at 
$T_{\text{c}} = (387\pm5)$~K and $T_{\text{s}} = (368\pm5)$~K. The inset shows normalised transmission at zero detuning as a function of magnetic field strength, where the black lines indicate the theoretical transmission while the dot points represent the experimental transmission values corresponding to the magnetic field strengths shown in the main plot. Residuals below indicate the agreement between experimental data and theoretical fits.}. 
\label{fig:spec with B}
\end{figure}

Figure \ref{fig:spec with B} illustrates the transmission spectra of the potassium D1 line measured at $T_{\text{c}} = (387\pm5)$~K and $T_{\text{s}} = (368\pm5)$~K for a 25~mm vapour cell containing 60\,Torr of neon buffer gas. The spectra are recorded at three different magnetic field strengths that are measured by fitting the spectra to $ElecSus$. The measured magnetic fields are: $(1200 \pm 5)$~G; $(970 \pm 4)$~G; and $(750 \pm 5)$~G. Magnetic field measurement errors were determined by varying the input light polarisation at a fixed magnetic field and fitting $ElecSus$ to extract the field for each polarisation state (see Section~\ref{sec:poldep}). Theoretical modelling of the magnetic field profile at these three permanent magnet separations (see Section~\ref{subsec:zeemandetails} for more details) predict axial root-mean-square magnetic field variations of 3\%, 13\%, and 23\% over the vapour cell length, respectively. Irrespective of this, the spectra agree well with $ElecSus$, as confirmed by the residuals. At lower magnetic field, the Zeeman splitting is small but clearly visible, with absorption peaks showing significant overlap. At B~\(\sim\)\,900~G the splitting is increased compared to B~\(\sim\)\,700~G, and the features are well-resolved. At the highest magnetic field, the spectrum shows the largest splitting, with the central dip becoming less pronounced. The broadening of the spectral lines is determined by Doppler (\(\sim\)\,800 MHz) and pressure broadening (\(\sim\)\,420 MHz) effects.

The inset panel shows the normalised transmission at zero detuning as a function of the applied magnetic field, where the black lines indicate the theoretical transmission while the dot points represent the experimental transmission values corresponding to the magnetic field strengths shown in the main plot. The observed increase in line-centre transmission reflects the increasing Zeeman splitting. The residuals in the lower panels illustrate a very good agreement between the experimental data and the theoretical fit. 

\subsubsection{Polarisation-dependent Zeeman spectra of potassium D1} \label{sec:poldep}

\begin{figure}[htbp]
\centering
\includegraphics[width=1.0\linewidth]{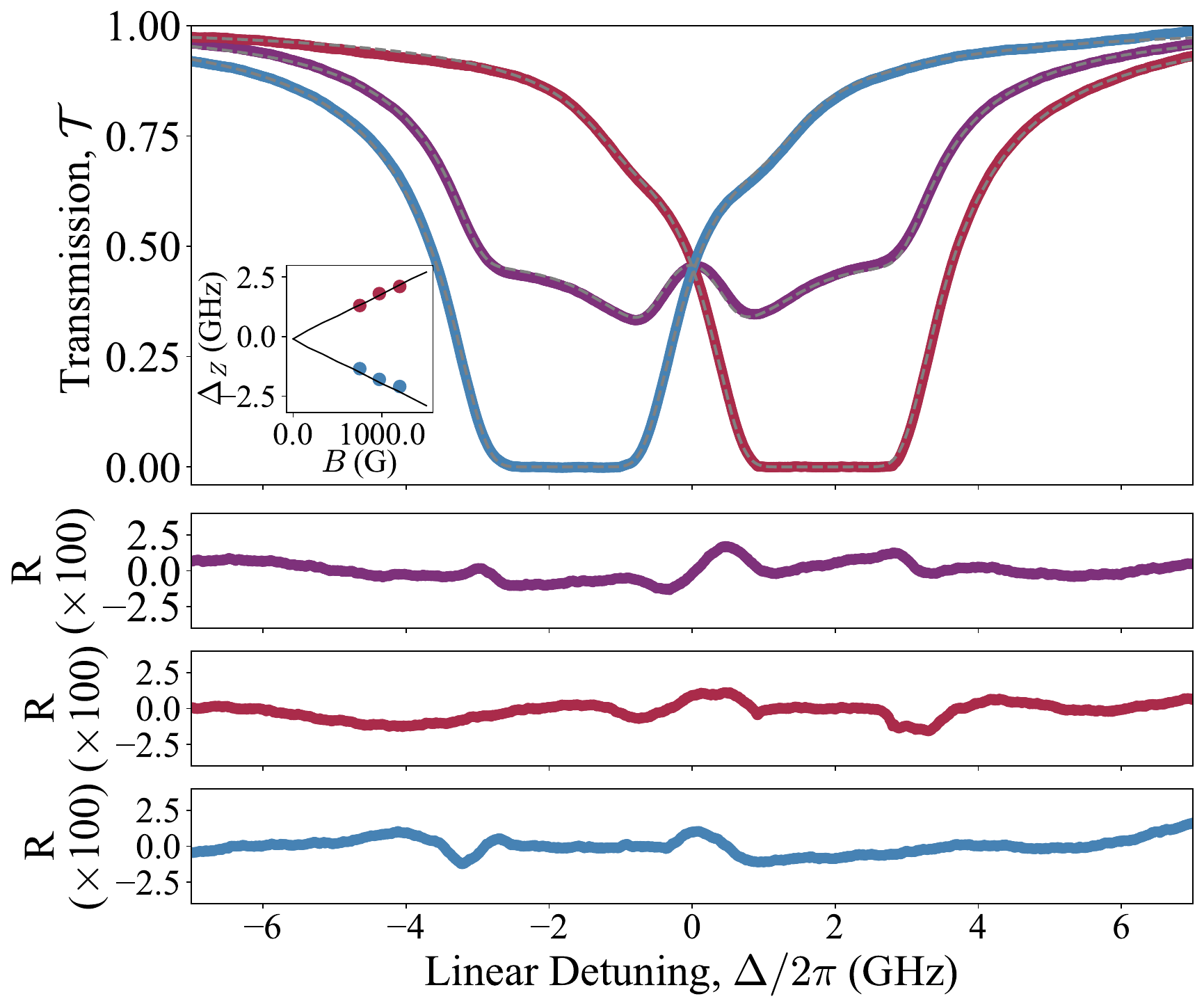}

\caption{Transmission spectra of the potassium D1 line at a magnetic field of $B~=~(970\pm5)$~G, measured with linearly polarised light (purple), left-circularly polarised light ($\sigma^{+}$, red), and right-circularly polarised light ($\sigma^{-}$, blue). The data were obtained using a 25 mm vapour cell with 60 Torr of neon at $T_{\text{c}}$ = $(408 \pm 5)$~K and $T_{\text{s}}$ = $(386 \pm 5)$~K. The inset shows the theoretical (black lines) and experiment detuning (frequency shift) of ($\sigma^{+}$, red dots) and ($\sigma^{-}$, blue dots) transitions for the three magnetic fields (1200$\,\pm\,5$)\,G, ($970\,\pm\,4$)\,G, and ($750\,\pm\,5$)\,G, confirming the linear Zeeman shift. Residuals below indicate the agreement between experimental data and theoretical fits.}. 
\label{fig:polarisation}
\end{figure}

Figure \ref{fig:polarisation} presents the transmission spectra of the potassium D1 line for three distinct input polarisations: linear horizontal (purple), left-hand circularly polarised (which drives $\sigma^{+}$ transitions, red), and right-hand circularly polarised (which drives $\sigma^{-}$ transitions, blue). The data were collected using a 25~mm potassium vapour cell containing 60 Torr neon buffer gas, maintained at a cell body temperature ($T_{\rm c}$) of (408~$\pm$~5)\,K and a stem temperature ($T_{\rm s}$) of (386~$\pm$~5)\, K. The magnetic field was measured to be $B$~=\,(970~$\pm$~4)~G by fitting the spectra to $ElecSus$.

~\\
For linearly polarised light, the spectrum exhibits two (nearly) symmetric absorption features, resulting from the combined contributions of both $\sigma^{+}$ and $\sigma^{-}$ transitions. In contrast, the $\sigma^{+}$ and $\sigma^{-}$ spectra display clearly separated absorption peaks, corresponding to the Zeeman-shifted transitions of the $m_{J}$ sublevels. These shifts are observed as negative and positive detunings for right- and left-hand circularly polarised light, respectively, illustrating the symmetry of the Zeeman effect. The spectra are not perfectly symmetric, as a consequence of there being more than one isotope. Notably, a small absorption dip near $\pm$1 GHz is visible, attributed to the $^{41}$K isotope.

It is possible to obtain complete extinction of the input light for the cases of left- and right -hand circularly polarised light.  In contrast, for linear light we can decompose the initial polarisation into equal admixture of left- and right -hand circularly polarised light.  As a consequence of the Zeeman effect when one hand of light couples strongly to the medium the other does not, and a minimum transmission of close to 50\% is obtained.

The inset depicts the experimental Zeeman detunings ($\Delta_{Z}$) of the $\sigma^{+}$ (red dots) and $\sigma^{-}$ (blue dots) transitions for three magnetic field strengths (1200 G, 970 G, and 750 G), alongside the corresponding theoretical predictions (black lines). In the high-field Paschen-Back (HPB) regime, the slope of the Zeeman shift for the potassium D1 line is expected to be $\pm\frac{3}{4}\mu_{\rm B}B$. For the magnetic fields used in this study, the Zeeman detunings ($\Delta_{Z}$) are $\pm$2.2, $\pm$1.7, $\pm$1.3 GHz, respectively, consistent with the predicted linear dependence. The residuals below the main plot demonstrate excellent agreement between the experimental data and the theoretical fits. 

This figure illustrates the strong influence of the polarisation state for transmission of light in the vicinity of the D1 transitions through a heated potassium vapour subject to an external magnetic field. To fully account for the transmission spectrum, especially in the wings, the buffer gas broadening has to be taken into account.

\section{Conclusion and Outlook}
\label{sec:conclusion}
This study investigated the spectral characteristics of the potassium D1 transition in the presence of neon buffer gas, focusing on the effects of pressure broadening, shift, and Zeeman shift in an external magnetic field. With a dual-temperature control system, we were able to measure the absorption spectra at various temperatures and buffer gas pressures which allowed us to obtain the values of broadening \((\Gamma)\) and shift \((S)\). Our findings confirm that the increased amount of buffer gas indeed leads to significant spectral broadening and a shift to lower frequencies. In addition, we measured the buffer gas filling pressures to be ($57.4\,\pm\,0.6$) Torr and ($96\,\pm\,1$) Torr, validating the bespoke request of 60\,Torr and 100\,Torr imposed on the vapour cell manufacturer.  The literature values for potassium-neon collisions of Pitz {\it et al.}~\cite{pitz2012pressure} were found to be in excellent agreement with our measurements. For the first time we were able to incorporate the buffer-gas shift and broadening into the modified Voigt profile via the {\it ElecSus} code, and found excellent agreement between the predicted and measured line profiles. Significant modifications are seen for cells with more buffer gas at elevated temperatures, especially in the wings of the profile. The theoretical prediction of the evolution of the Voigt profile from Gaussian- to Lorentzian-dominated was successfully reproduced in the experiment.

In the second part of this study, we explored the impact of an external magnetic field on the pressure-broadened potassium D1 spectrum using a 25 mm vapour cell. We obtained the magnetic field strength and analysed the Zeeman splitting in the hyperfine Paschen-Back regime by fitting the transmission spectra for linearly and circularly polarised components (\(\sigma^+\) and \(\sigma^-\)) to $ElecSus$. The measured shifts of the \(\sigma ^+\) and \(\sigma^-\) components showed a linear dependence on the applied field, which is consistent with the expected Zeeman effect for this regime. The excellent agreement between experiment and theory indicates the successful inclusion of Zeeman effect, and collisional broadening and shift into our model for alkali-metal vapour transmission. 

While motivated by design of magneto-optical filters in solar physics, the results of this study further enhance our understanding of potassium spectroscopy as well as the interactions between alkali-metal atoms and buffer gases. The insight gained here into the evolution of the lineshape will also allow optimisation of other sensing techniques that utilise potassium vapours~\cite{schwartz2023laser, ZHANG2001147, Hoffner:05, Kudenov:20, Fricke-Begemann:02}.

As noted above for potassium the linewidth of the transition exceeds the atomic hyperfine splitting leading to one broad transition. Future work could study heavier alkali metals, such as rubidium and caesium, with a range of buffer gases to provide a further test of the model of alkali-metal vapour transmission encapsulated in the {\it ElecSus} code.\\

\noindent\textbf{Data availability statement}
The data that support the findings of this study are openly
available at the following URL/DOI: 10.15128/r2kw52j810w. Data underlying the results presented in this paper are available from~\cite{data}.\\

\noindent\textbf{Acknowledgements.} 
The authors thank Jack D. Briscoe for valuable $ElecSus$ discussions. \\

\noindent\textbf{Funding.} IGH acknowledges the funding received from EPSRC (Grant No. EP/R002061/1) and the UK Space Agency (Grant No. UKSAG22\_0031\_ETP2-035). SAA acknowledges support from Najran University, Najran, KSA (Grant No. 443-16-4151).\\
\

\noindent\textbf{Disclosures.} The authors declare no conflicts of interest.

\bibliographystyle{unsrt}
\bibliography{references}


\end{document}


\title{Supplemental Material: The Role of Buffer Gas in Shaping the D1 Line Spectrum 
of Potassium Vapour
}

\author{Sharaa A. Alqarni, Danielle Pizzey, Steven A Wrathmall, and Ifan G Hughes}

\affiliation{Department of Physics, Durham University, South Road, Durham, DH1 3LE, UK}

\affiliation{Corresponding author: sharaa.alqarni@durham.ac.uk}

\maketitle

\begin{figure}[htbp]
\centering
\includegraphics[width=\linewidth]{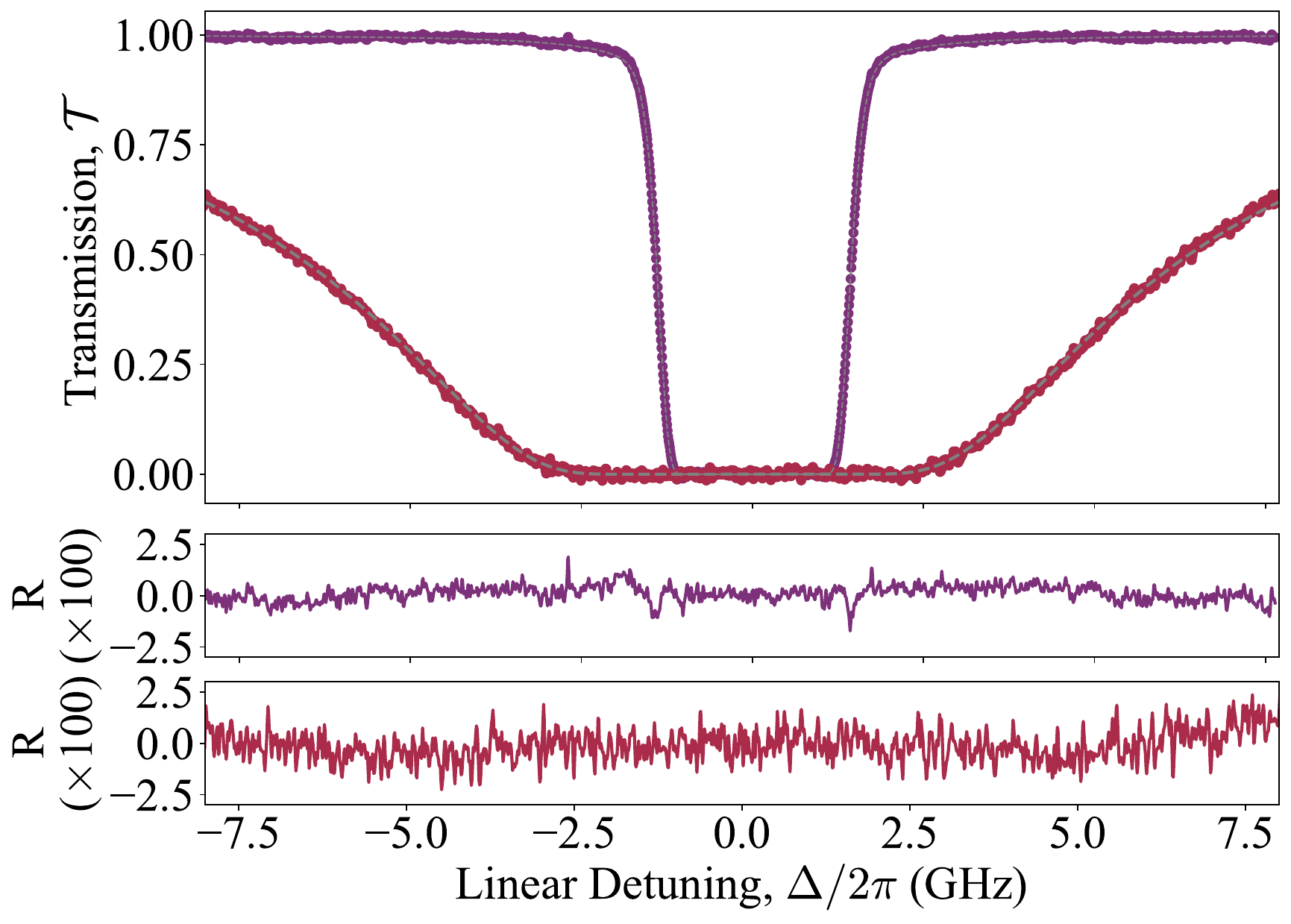}
\caption{Transmission plot for the comparison between experiment and theory for the K D1 line, through a 75 mm vapour cell as a function of linear detuning, $\Delta /2\pi$. Solid (purple and red) and dotted (grey) lines show measured and expected transmission for ${T_{\mathrm{c}}}$= \SI[separate-uncertainty]{420\pm5}{k}
and ${T_{\mathrm{s}}}$ = \SI[separate-uncertainty]{400\pm5}{K} of cell without buffer gas and 100 Torr cell, respectively. The transmission for 100 Torr cell has $\Gamma$ = 
\SI[separate-uncertainty]{765\pm2}{MHz}
and $S$ = \SI[separate-uncertainty]{-158\pm1}{MHz}. The inset displays the predicted transmission of 100 torr cell at very large detuning. Below the main figure is plot of the residuals.}. 
\label{fig:100 torr with theory}
\end{figure}

Figure~\ref{fig:100 torr with theory} compares the experimental transmission spectra with theoretical predictions for 75~mm potassium vapour cells with no buffer gas and 100 Torr neon, respectively, at $T_{\rm c}$ = \SI[separate-uncertainty]{420\pm5}{K} and $T_{\rm s}$ = \SI[separate-uncertainty]{400\pm5}{K}. The main plot demonstrates the characteristic absorption feature of the potassium D1 line, with transmission approaching zero at the line centre and gradually recovering to near 100\% at detunings around 1 GHz for the buffer gas free spectrum. The 100 Torr neon spectrum clearly exhibits pressure-induced effects, with a measured broadening ($\Gamma$) of \SI[separate-uncertainty]{765\pm2}{MHz} and a shift ($S$) of \SI[separate-uncertainty]{-158\pm1}{MHz}. Our 16 GHz scanning range was insufficient to capture the full spectral profile; from $ElecSus$ the transmission returns to 100\% at detunings exceeding 50 GHz.

\bibliographystyle{unsrt}
\bibliography{references}
